\documentstyle[12pt, fleqn]{article}
\def\ref{\par\noindent\hangindent=6mm\hangafter=1}
\begin{document}
\baselineskip 8mm

\begin{center}

{\bf FeII and FeI emission in IRAS 07598+6508 and PHL 1092}

\bigskip
\vspace{10mm}

John Kwan$^1$, Fu-zhen Cheng$^2$, Li-Zhi Fang$^3$, Wei Zheng$^4$
and Jian Ge$^5$

\bigskip

\end{center}

1. Department of Physics and Astronomy, University of Massachusetts,
Amherst, MA 01003

2. STScT, 3700 San Martin D., Baltimore, MD 21218

3. Department of Physics, University of Arizona, Tucson, AZ 85721

4. Department of Physics and Astronomy, Johns Hopkins University,
Baltimore, MD 21218

5. Steward Observatory, University of Arizona, Tucson, AZ 85721

\newpage

{\bf Abstract}

\bigskip

One of the puzzles in understanding the spectra of active galactic
nuclei (AGN) is the origin of the FeII emission. FeI emission,
if present, will help reveal the physical conditions of the emitting
gas. In an attempt to verify the presence of FeI lines, high S/N
spectra of two FeII-strong quasars, IRAS 07598-6508 and PHL 1092,
were obtained at the Multiple Mirror Telescope and the Steward 2.3 m
Telescope. We have identified emission lines of FeI and TiII. The
source of energy for FeII, FeI and TiII emission is probably not
from ionization by the photon continuum, but heat. The high rate
of energy generation and the presence of both high and low velocity
gas indicate that the heat is generated not over a large area, but
a narrow band in accretion disk, in which the rotational speed
decreases rapidly.

\newpage

\noindent{\bf 1. Introduction}

\bigskip

One of the puzzles in understanding the spectra of active galactic nuclei
(AGN) is the origin of the FeII emission. This was emphasized by Wills,
Netzer, \& Wills (1985) who pointed out that in many AGN the strong FeII
emission, in relation to that of Ly$\alpha$, cannot be accounted for by
standard photoionization models (Kwan \& Krolik 1981; Kwan 1984;
Wills, Netzer, \& Wills 1985; Collin-Souffrin, Hameury, \& Joly 1988).
Adding to the bewilderment is the discovery of several AGN with super
strong FeII emission (Bergeron \& Kunth 1980; Lawrence et al. 1988;
Lipari, Terlevich, \& Macchetto 1993). The FeII $\lambda$4570/H$\beta$
ratios in them appear to be several times larger than those found in
strong FeII emitters. Bergeron \& Kunth (1980) also noted the presence
of FeI features in PHL 1092, and Wills, Netzer, \& Wills (1985)
mentioned that in some objects they found a flux excess above their
model fits at wavelengths corresponding to those of strong FeI multiplets,
but cautioned that FeII lines might actually be the contributors. FeI
emission, if present, will help reveal the physical conditions of
the emitting gas, but, as far as we know, there has been no further
published record.

In an attempt to understand the FeII emission, and to verify the presence
of FeI lines, we have obtained spectra of IRAS 07598+6508 and PHL 1092,
using the MMT telescope. In the next section we give details of the
observations. We present arguments for the identification of FeI and TiII
features in \S 3, and discuss the origin of the FeII, FeI and TiII emission
in \S 4.

\bigskip

\noindent{\bf 2. Observations}

\bigskip

One set of observations of IRAS 07598+6508 was made at the MMT
Spectrograph with the T I 1200$\times$800 CCD in the red channel on
November 3, 1992. A 600 lines mm$^{-1}$ grating was used. The wavelength
coverage was between 3500 and 5250 \AA, with a resolution of about 5 \AA.
The slit width was 3.5". Two exposures, 20 minutes each, were taken.
The seeing was poor, so the spectrum was not photometric. Another set of
spectra of IRAS 07598+6508 was obtained at Steward Observatory 2.3m
telescope
with a B \& C spectrograph and a 1200$\times$800 CCD on February 2, 1994.
Three exposures of a total of 100 minutes were taken with a slit width of
3" and a 400 lines mm$^{-1}$ grating in the second order, covering
a wavelength range of 3600 to 5200 \AA. The flux level of this spectrum is
accurate to within 15\%. The redshift of this object is 0.1483, as derived
by Lawrence et al. (1988).

The observations of PHL 1092 were made at the Cassegrain focus of the
Steward Observatory 2.3m telescope with a B \& C spectrograph and a
1200$\times$800 CCD on September 23, 1992. A 600 lines mm$^{-1}$ grating
was used. The wavelength coverage was between 4050 and 6150 \AA. The width
of
the long slit was 1.5", and three exposures, each 40 minutes long, were
taken
sequentially. PHL 1092 has a redshift of 0.396 (Schmidt 1974).

Standard data reduction was performed using IRAF. The signal to noise ratio
of each spectrum is about 50. While the continuum level is difficult to
determine, we generate a pseudo-continuum by fitting several points where
the observed fluxes are low, and plot the ratio of the observed flux to the
pseudo-continuum flux.  The pseudo-continuum fit is made with a $n=2$
polynomial. In this way neighboring features in the spectrum become more
distinct because of the smaller vertical scale. The spectra of IRAS
07598+6508 and PHL 1092, when shifted to rest-frame wavelengths, are very
similar. We will use the former spectrum to demonstrate our identification
of the spectral features.

\bigskip

\noindent{\bf 3. Identification of FeI and TiII Features}

\bigskip

Fig.1 shows the normalized spectrum of IRAS 07598+6508 from all
our observations. In the space above the spectrum we have marked
the positions of FeII lines, together with their multiplet numbers.
These lines arise from the six odd-parity terms that lie between 4.76 and
5.91 eV above the ground state. Their placements in Fig.1 are such
that multiplets at same height originate from the same upper term.
In order of increasing height, the six vertical levels where the
FeII multiplets are marked above the spectrum correspond
to the upper terms that are designated z$^6$P,
z$^6$F, z$^6$D, z$^4$D, z$^4$P, and z$^4$F respectively. For example,
multiplets 16, 23, and 29 originate from z$^4$P.

Collisional excitation of the above six terms, followed by permitted
radiative decays to metastable even-parity terms, is a major source of
FeII emission. The next aggregate of odd-parity terms lies significantly
higher, more than 7.4 eV above the ground state. The marked positions in
Fig. 1 are then where FeII emission is expected to occur within the
observed wavelength range of 3600-5200 \AA. The vertical length of each
marked line depends on the product g$_u$A, where g$_u$ is the degeneracy of
the upper state and A is the spontaneous emission rate. It is equal to
[0.002 + 0.01 log(g$_u$A/10$^{4}$)], for 10$^4$ $\leq$ g$_u$A $\leq$ 10$^7$.
Lines with g$_u$ $<$ 10$^4$ are not plotted, and lines with g$_u$A $>$ 10$^7$
have a maximum vertical length of 0.032. This arrangement provides some
distinction between strong and weak lines. The relevant atomic
data for FeII, and later on for FeI and TiII are gathered from Fuhr, Martin,
\& Wiese (1988), and Martin, Fuhr, \& Wiese (1988) when available, and from
Kurucz (1974, 1981), and Kurucz \& Peytremann (1975) otherwise.

Even a casual look at the observed spectrum shows that many emission features
can be readily identified as FeII lines. Examples are: the series of peaks
at 4800-5000 \AA that nicely coincide with the positions of strong lines from
multiplets 27 and 28; the broad feature at 3600-3800 \AA \ that is most
likely
contributed by multiplets 1, 6 and 7.

We identify, however, three spectral regions in Fig.1 whose features cannot
be accounted for by FeII lines. The first region is from 4050 to 4700 \AA.
There several emission peaks are present that have no coincident  FeII lines.
The several FeII multiplets that are present, except for number 3, also
cannot be responsible for the bulk of the emission at their positions. Thus,
multiplets 23 and 29 are expected to be weak because their upper term has
a faster decay route in multiplet 74 (observed wavelength at 7060-7413 \AA).
The spontaneous emission rate via multiplet 74 is at least three times larger
than that of any line of multiplet 23 or 29. Optical depth effects will not
be strong as the lower term of multiplet 74 is not only $\sim$ 3.9 eV above
the ground state but also $\sim$ 1.3 eV above the lower terms of multiplets
23 and 29. In the same way multiplet 14 is much weaker than multiplets 27
and 38. It is also positioned incorrectly in that its strongest line occurs
close to a dip in the observed spectrum. The single marked line belonging
to multiplet 15 is extremely weak as its upper state can spontaneously decay
via a transition of multiplet 49 (observed wavelength at 6011 \AA) at a rate
that is $\sim$20 times faster. Multiplet 3, in spite of small A rates, can
become strong when multiplet 1 and the UV multiplets of z$^6$D become very
optically thick. Together with the CaII H and K lines, it may contribute to
the two emission features at 4520 and 4560 \AA. For the other features
between 4050 and 4700 \AA, we conclude that they are not result of emission
from the six lowest odd-parity terms.

The second spectral region we identify where there are few marked FeII lines
is  from 3810 to 3960 \AA. Multiplet 5 and two lines of multiplet 16 lie
in that spectral range. Like multiplets 23 and 29, multiplet 16 does not
provide a fast decay route for their upper term. The spontaneous emission
rate of its line marked at 3955 \AA \ is only $\sim$ 1/8 of the total rate
from the same upper state down to all levels above a$^2$P, the lower term of
multiplet 16. Its line marked at 3923 \AA \ is stronger, having an A rate
about
the same as its corresponding total rate to all levels above a$^2$P. A
similar situation hold for multiplet 5. Its upper term, z$^6$P, decays to
a$^6$S (multiplet 42 at observed wavelengths of $\sim$ 5800 \AA), which lies
$\sim$  1.2 eV above the lower term of multiplet 5, at $\sim$ 300 times
faster. Based on these considerations, we conclude that
the broad feature between FeII multiplets 1 and 4 has a different origin.

The third spectral region is from 5050 to 5150 \AA. Of the four marked lines
present, only one, at 5072 \AA \ (from multiplet 27) competes effectively
for the de-excitation of its upper state. Each of the other three has a
spontaneous emission rate that is more than a factor of 10 smaller than the
decay rate to a level above its lower term. Although not as evident as is the
case with the first or second spectral region, additional emission is also
needed here.

We have, thus far, pointed out spectral features in Fig.1 that are not being
emitted by the six lowest odd-parity terms of FeII. Examining the permitted
transitions  of more highly excited terms, we are also confident they are
not responsible for the bulk of the emission in those features. First,
collisional excitation of those terms require $\sim$ 2 eV or more energy
above that needed to excite the six lowest odd-parity terms. Then those lines
that lie within the three spectral regions all have A values of $\leq$ 10$^6$
s$^{-1}$, and are not the dominant transitions from their respective upper
states. It is difficult to see how collisional excitation to the higher
odd-parity terms can produce the features in the three spectral regions
which,
taken together, comprise an amount of emission, within the observed
wavelength
range of 3600-5200 \AA, that is comparable to the total in the identified
lines from the six lowest odd-parity terms. Second, there are not enough
strong lines, with line strength values gf $\geq$ 10$^{-3}$, from upper
states
within 9eV from the ground state to account for all the features in the
three spectral regions.

FeII line coincidence, due to wavelength proximity to within a few Doppler
widths, between a UV line originating from one of the six lowest odd-parity
terms and another originating from an upper term may over populate a highly
excited state and lead to a stronger emission of UV and optical lines from
that state. We have examined whether the emission features in the three
spectral regions originate from this fluorescence process by employing
the list, tabulated by Netzer and Wills (1983), of line pairs whose
wavelength difference corresponds to a velocity shift of $\leq$ 7.5 km
s$^{-1}$. From that list we identify about 20 line pairs such that the
upper state that may be over populated as a result of a line coincidence
has a
transition lying within the three spectral regions. Comparing the spontaneous
emission rate of that transition with those of others from the same upper
state, we find that the former rate is generally less than 10$^{-3}$ of the
total rate. The fluorescence process will therefore the enhance little the
emission in the optical transitions which the three spectral
regions encompass. Rather, it will enhance primarily the emission in the
many strong UV transitions. Radiative trappings of these transitions, whose
lower states lie at several eVs above the ground state, will be less severe
than those of UV transitions originating from the six lowest odd-parity
terms.
There are also features in the three spectral regions that cannot be
identified with transitions from the special upper states. Thus, while
we cannot rule out a small contribution to a few features in the three
spectral
regions from FeII line coincidence and fluorescence, we are confident this
process does not produce the bulk of those emission features.

We believe emission from ions other than FeII is required. In the space
below
the spectrum in Fig.1 we have marked the positions of emission produced by
FeI and TiII. The TiII lines are marked on the vertical level of 0.99 and
their
 multiplet numbers are prefixed by Ti. The many more FeI lines are marked
farther below, identified by only their multiplet numbers.

The FeI multiplets shown originate from the 15 lowest odd-parity terms.
Unlike those of FeII, the odd-parity terms of FeI are distributed fairly
uniformly in an energy-level diagram, so there is not an obvious grouping
of them, as in the case of FeII.

Many of the FeI lines marked represent dominant decay routes of their
respective upper states. Their spontaneous emission rates, however, occupy a
very wide range, from $\sim$ 10$^3$ to $\sim$ 10$^8$ s$^{-1}$, so the
procedure used earlier to signal the g$_u$A value of a line by the plotted
 vertical length will not provide a fair representation of dominant
transitions that happen to have very low A values. For the FeI lines,
then, the following procedure is adopted. The vertical length is equal to
[0.1(g$_i$/$\Sigma_i$g$_i$)(A$_{ij}$/$\Sigma_{ij}$A$_{ij,k}$)], where
g$_i$ is the degeneracy of the upper state, $\Sigma_i$ g$_i$ is the total
degeneracy of the upper term, A$_{ij}$ is the spontaneous emission rate of
the
line, and $\Sigma_{ij}$ A$_{ij,k}$ is the total spontaneous emission rate
from the upper state to all lower states j of all lower terms k. We assume
that population into a given state i of an upper term is proportional to
g$_i$, so the vertical length represents the probability of spontaneous
emission of the line photon upon collisional excitation of the upper term.
Only lines with vertical lengths equal to 0.001 or more are plotted.

The large number of FeI terms involved and the limited available space in
Fig.1 does not allow us to place the line markings in the orderly fashion
that was done in the FeII case. Of the FeI multiplets indicated, only
several
share the same upper term. They are multiplets 7 and 18 which have z$^3$F as
their upper term, multiplets 23 and 41 which have z$^5$G as their upper
term, and multiplets 24 and 42 which have z$^3$G as their upper term.

A casual comparison between the FeI line positions and the emission features
in the first spectral region mentioned above shows that many of the
observed
features coincide with strong FeI lines. A firm identification must await
detailed calculations of the line intensities. Unfortunately, reliable
collisional cross-sections are needed. They are not available, and our
qualitative discussion of the FeI excitation below is accordingly hampered.

The three lowest odd-parity terms of FeI are z$^7$D, z$^7$F, and z$^7$P.
They
lie from 2.4 to 3.0 eV above the ground state. The dominant decay routes from
them are contained in multiplet 1 (which lies outside the observed spectrum),
2, and 3 respectively. Judging from Fig.1, we find that the emissions that
can
be identified with multiplets 2 and 3 are fairly weak, particularly so in
the
case of multiplet 3. Thus, despite the proximity of those two terms to the
ground state, population into them must be relatively slow if other FeI
multiplets are to be identified with more prominent features. a plausible
cause is that multiplets 2 and 3, linking their respective upper terms to
the
ground term a$^5$D, have very small oscillator strengths (f $\sim$
4$\times$10$^{-5}$, so collisional excitation of them may be comparatively
weak.

The next group of odd-parity terms are z$^5$D, z$^5$F, and z$^5$P, lying
from 3.2 to 3.7 eV above the ground state. The dominant decay routes from
them are contained in multiplets 4, 5 and 6 respectively, which connect down
to the ground term. They have large oscillator strengths
(f $\sim$ 10$^{-2}$).
When their optical depths exceed about 10, 200, and 20 respectively,
multiplets 15, 16, and 60, which have observed wavelengths
longward of 5200\AA,
will compete for the de-excitation of z$^5$D, z$^5$F, and z$^5$P
respectively.
Until the FeI column density becomes very high then, we expect that
population into these three excited terms will produce emission largely in
multiplets 4, 5 and 6.

Most of the multiplets that contribute to the prominent features in the
first spectral region originate from the group of odd-parity terms
comprising y$^5$D, y$^5$F, z$^5$G, z$^3$G and y$^3$F. These five excited
terms lie from 4.1 to 4.65 eV above the ground state. They, together with
y$^3$D (lying at $\sim$ 4.8 eV), have the distinction that their outer-shell
electronic configuration, 3d$^7$ (a$^4$F) 4p, has the same inner structure
as that, 3d$^7$ (a$^4$F) 4s, of the two lowest excited terms a$^5$F and
a$^3$F. Thus, radiative transition coefficients between those odd-parity
terms and one or both of the two lower even-parity ones are particularly
large. Multiplets 20 (y$^5$D-a$^5$F), 21 (y$^5$F - a$^5$F), 23 (z$^5$G -
a$^5$F), 41 (z$^5$G - a$^3$F), 24 (z$^3$G - a$^5$F), 42 (z$^3$G - a$^3$F),
and 43 (y$^3$F - a$^3$F) all have transitions with very strong oscillator
strengths (f $\sim$ 0.1). We judge that the majority of the population
into those five excited terms will produce emission in the above-mentioned
multiplets. To generate the prominent features in the first spectral region
then, a strong excitation of those five terms, despite their relatively
high excitation energies, is required.

FeI Multiplets 2 and 41 can contribute to the emission features in the
third spectral region mentioned above, but the broad feature in the second
spectral region remains to be identified. Two ions, Cr II and TiII, have
strong emission lines there. If the elemental abundance is solar-like, Cr is
5.2 times as abundant as Ti. But collisional excitation of the appropriate
Cr II terms, which lie at $\sim$ 6.14 eV above the ground state, requires
$\sim$ 2.36 eV more energy than that needed to produce the Ti II lines. This
latter consideration, and our judgment that the TiII lines are better
positioned to match the broad feature lead us to decide on TiII.

The TiII lines that are marked in Fig.1 originate from the two lowest
odd-parity terms. Multiplets 1, 6, and 11 have z$^4$G as their upper
term which lies at $\sim$ 3.68 eV from the ground state, while multiplets
2, 7 and 12 have z$^4$F as their upper term which lies $\sim$ 0.16 eV farther
above.  We expect them to be less optically thick than the FeII multiplets,
so only the stronger lines, with values of g$_u$A $\geq$ 3$\times$
10$^6$ s$^{-1}$, have been plotted. The vertical length of a marked TiII
line is equal to [0.002 + 0.01 log(g$_u$A/3$\times$10$^6$)], with a maximum
of 0.032.

Whereas FeII is a likely cooling agent of the gas, both FeI and TiII are
not. The former is readily ionized to FeII, while the latter has an
abundance,
assuming solar-like, that is no more than 1/300 that of Fe. A condition that
favors collisional excitation of the FeI and TiII lines is the comparatively
low energy needed. Thus it takes $\sim$ 1.9 eV less energy to produce TiII
multiplet 1 than to produce FeII multiplet 27. This translates, at a
temperature of 7000 K, to a factor of $\sim$ 23 in the excitation rate. When
the gas temperature is low, and the ionization condition is such that a small
fraction ($\geq$ 0.01) of Fe is FeI, cooling due to collisional excitation
of TiII and FeI will become important.

Fig.2 shows the observed spectrum of IRAS 07598+6508 obtained on February
2, 1994, and the observed spectrum of PHL 1092. The two spectra are shift
to rest-frame wavelengths so that they can be compared directly. We have
not labelled the emission features, but it is clear that the two spectra
appear very similar. The first and second spectral regions, at rest-frame
wavelengths of 3520-4100 \AA, and 3320-4350 \AA respectively, can be
readily identified.

\bigskip

\noindent{\bf 4. Discussion}

\bigskip

The above identification of FeI and TiII features, if correct, indicates a
small but significant amount of FeI. A quantitative determination of the
[FeI]/[FeII] abundance ratio is difficult at present because the underlying
continuum is not firmly established, so the observed line intensities are
uncertain, and reliable FeII and FeI collisional cross-sections are not
available. We guess it is probably greater than 0.01. We can make a somewhat
better estimate of the column density, as follows. The line $\lambda$4233.17
(observed at 4860.95 \AA) of FeII multiplet 27 (b$^4$P-z$^4$D) is strong.
Another line from the same upper state, $\lambda$4731.44 (observed at
5433.11 \AA), belonging to multiplet 43 (a$^6$S - z$^4$D) and having a
spontaneous emission rate that is 30 times smaller, is much weaker as we
judge from the spectrum of IRAS 07598+5608 obtained by Lawrence et al.
(1988).
The $\lambda$4233.17 opacity is then less than or about 30. Assuming an
absence of a velocity gradient within the emission column, the Fe II column
density is N$_{FeII} \  \leq$ 2 $\times$ 10$^19$ (7000 K/T)$^{3.6}$
($\Delta$ v/20 km s$^{-1}$) cm$^{-2}$, where $\Delta$v is the intrinsic
linewidth (due to thermal or turbulence broadening) and the gas temperature
T is assumed to lie between 6000 and 8000K.

We can also make a lower estimate of the luminosity of the FeII, FeI and
TiII
emission. Assuming that the underlying continuum is smooth, we use the low
points (at observed wavelengths of 3615, 4245, and 4690 \AA) of the spectrum
obtained on February 2, 1994 to demarcate an upper bound to the underlying
continuum. It can be roughly gauged from Fig. 2 that over the observed
wavelength range of 3600-5200 \AA the emission features altogether
constitute
a luminosity that is about 1/7 or more of the continuum luminosity.
assuming an H$_0$ of 50 km s$^{-1}$ Mpc$^{-1}$, an $\Omega$ of 1, we
calculate this luminosity of the emission features to be
4.2$\times$10$^{10}$ L$_{\odot}$. It can also be seen from the spectrum of
Lawrence et al. (1988) that over the observed wavelength range of 5000-9000
\AA the emission features there, excluding H$_{\alpha}$, H$_{\beta}$, and He
I $\lambda$5876, and which consist primarily of FeI and FeI multiplets, are
comparably strong in relation to the underlying continuum. FeII UV
multiplets, at rest wavelengths from 2200 to 3000 \AA, are expected to be
even stronger than FeII optical multiplets. Thus, over the observed
wavelength range of 2500-9000 \AA the luminosity in the FeII, FeI and
TiII features is a significant fraction of the total.

Another item of information indicated by the observed spectrum is the
narrowness of the lines. This has been noted by Bergeron \& Kunth (1980)
from their spectrum of PHL 1092. Consider the FeII $\lambda$4233.17 line,
for example. While the linewidth at zero intensity is not easily determined
because of line blending and the uncertain continuum level, the narrowness
at the peak clearly stands out. When the close proximity of neighboring
lines is taken into account, other emission features, such as the ones at
 observed wavelengths of 4315 and 4795 \AA, also point to sharply-peaked
individual lines. The observed width of a line is generally ascribed to
Doppler broadening, so the narrow line peak indicates that some of the
emitting gas have bulk velocities less than 500 km s$^{-1}$.

We do not think the source of energy for the very strong FeII, FeI, and
TiII emission is ionization by the photon continuum. The FeII column density
estimated above indicates, for solar-like abundances, a nucleon column
density
that is sufficient to absorb only the UV ($\geq$ 13.6 eV) and soft x-ray
 continuum. Photoionization calculations at that column density and with the
typical continuum energy distribution will not be able to produce the strong
 FeII and FeI luminosity estimated earlier in relation to the underlying
continuum. They will also not be able to produce the relative intensity
between FeII and hydrogen emission. Lawrence et al. (1988) estimate FeII
4570/H$_{\beta}$ to be in the range of 4-8, about an order of magnitude
higher
than the typical value in AGNs with strong FeII emission. It can also be
seen from Fig. 1 that H$_{\gamma}$, and H$_{\delta}$, at observed
wavelength
of 4984 and 4710 \AA respectively, are almost invisible amid the FeII and
FeI features. It is probably impossible to produce a luminosity in FeII and
FeI lines that is at least several times the luminosity in the hydrogen
Balmer lines via absorption of the photo continuum beyond 13.6 eV, since
photoionization of hydrogen necessarily converts a large fraction of the
continuum energy into hydrogen quanta. Enhancement of the FeII and FeI
emission by choosing the nucleon density such that continuum from 7.78 to
11.26 eV were also absorbed within the same column density as result of
photoionization of FeI would not work, as the FeI luminosity would then
be stronger than the FeII luminosity, which is not the case.

We believe the source of energy for the FeII and FeI luminosity is most
likely heat. If heat is generated at a rate such that the equilibrium
temperature is less than 8000 K, collisional excitation of FeII and FeI
multiplets will be more effective coolants than collisional excitation of
the Balmer lines of hydrogen, largely because higher excitation energies are
needed in the latter process. The very high FeII and FeI luminosity
estimated earlier suggests to us that the source of heat is probably not
derived from stars, since heating generated from the interaction of HII
regions and supernova explosions with the interstellar medium does not
generally liberate a large fraction of its energy at temperatures of
less than 8000 K.

We think it is simpler and more efficient to produce the heating  from
an accretion disk around a massive black hole. The high rate of energy
generation and the presence of both high and low velocity gas indicate
to us that the heat is generated not over a large area such as the Keplerian
dependence of the rotational speed on radius produces the velocity
dispersion.
This is because the high and low velocity gases would then lie at very
different distances from the black hole, and it would be difficult to
generate a heating rate such that FeII emission is the dominant coolant over
the wide range of distance. We speculate that the heat may be generated in a
narrow band in which the rotational speed decreases rapidly, in a way
described below.

Recently, Hirotani et al. (1992) have studied the interaction between
accreting matter and magnetic field in the magnetosphere of a Kerr
black hole.
In their model the matter originates from a geometrically thin disk, and
the accretion onto the black hole occurs along the magnetic field lines
which
arise from the disk and thread the event horizon. The angular velocity
$\Omega_F$ of such a field line will depend on $\Omega_H$, the angular
velocity of the black hole, and $\Omega_K(r)$, the angular velocity of
the rotating disk. Studies of time-stationary, axisymmetric flow in the
magnetosphere of a Kerr black hole find that $\Omega_f \leq \Omega_H$
(Phinney 1983; Punsly \& Coroniti 1990). At the places where matter climb
onto the field lines, which depend on the magnetic field strength, it
is likely that $\Omega_k > \Omega_F$, and heat will be generated as the
matter loses part of its angular momentum before falling along the field
lines.
This situation is analogous to what Ghosh \& Lamb (1979) envision in their
model of accretion onto a neutron star from the surrounding disk.

We think that the FeII emission is produced in the region mentioned above
where orbital energy is dissipated. In the case of IRAS 07598+6508
we estimate that the rotational speed decreases from $\sim$ 5$\times$10$^3$
km s$^{-1}$ to 5$\times$10$^2$ km s$^{-1}$. Taking account of the large
velocity gradient within the dissipation region, the FeII column density
in the direction perpendicular to the disk is then N$_{FeII}$ $\leq$
2$\times$10$^{21}$ (7000 K/T)$^{3.6}$ cm$^{-2}$. If the energy dissipation
occurs at T $<$ 800 K, FeII emission is most probably the major coolant. At
higher temperatures hydrogen line and continuum emission will begin to
dominate. The weak hydrogen emission, in relation to FeII emission, that is
observed and the conspicuous presence of FeI and TiII features, which
require
lower excitation energies than the FeII features, are then mutually
consistent.

Once the gases radiate away their excess orbital energy, they will corotate
with the field lines and fall towards the black hole. The bulk of their
gravitational potential energy is likely to be released in the magnetic
funnel close to the event horizon, thereby generating the major portion of
the continuum energy. Away from this continuum source, the diffuse accreting
gas will probably be heated to the Compton temperature (Krolik, McKee \&
Tarter 1981; Mathews \& Ferland 1987). If dense, much cooler condensations
can be formed within the flow, they will be logical candidates for the broad
emission-line clouds. With their infalling velocity in creasing rapidly
from 0 to $>$ 10$^4$ km s$^{-1}$, the line intensities and profiles they
produce may match the observed results more easily than kinematic models in
which the high and low velocity emission occur at very different distances.

While our speculation of the origin of the FeII emission is stimulated
by its unusual strength in IRAS 07598+6508 and PHL 1092, we suspect the
same origin is true for those AGN whose strong FeII emission cannot be
accounted for by photoionization models. It would be useful to examine the
FeII emission of these objects at high signal to noise and with careful
modeling to see if the physical conditions and kinematic structures
required point to a similar scenario.

The observations reported here were obtained with the MMT, a facility
operated by the University of Arizona and SAO. We acknowledge P. Smith and
MMT staff for their assistance. F. Cheng thanks STScI for financial support
and much help on data reduction.

\newpage

\noindent{\bf References}

\bigskip

\ref Bergeron, J., \& Kunth, D. 1980, A\&A, 85, L11

\ref Collin-Souffrin, S., Hameury, J., \& Joly, M. 1988, A\&A, 205, 19

\ref Fuhr, J.R., Martin, G.A., \& Wiese, W.L. 1988, J. Phys. Chem.
Ref. Data, 17, Suppl. 4

\ref Ghosh, P., \& Lamb, F.K. 1979, ApJ, 232, 259

\ref Hirotani, K., Takahashi, M., Nitta, S,, \& Tomimatsu, A. 1992,
ApJ, 386, 455

\ref Krolik, J.H., McKee, C.F., \& Tarter, C.B. 1981, ApJ, 249, 422

\ref Kurucz, R.L. 1974, Smithsonian Ap. Obs. Spec. Rept, No. 359

\ref Kurucz R.L. 1981, Smithsonian Ap. Obs. Spec. Rept, No. 390

\ref Kurucz, R.L., \& Peytremann, E. 1975, Smithsonian Ap. Pbs. Spec.
Rept.No. 362

\ref Kwan, J. 1984, ApJ, 283, 70

\ref Kwan, J. \& Krolik, J.H. 1981, ApJ, 250, 478

\ref Lawrence, A., Saunders, W., Rowan-Robinson, M., Crawford, J.,
Ellis, R.S.,
Frenk, C.S., Efstathious, G., \& Kaiser, N. 1988, MNRAS, 235, 261

\ref Lipari, S.., Terlevich, R., \& Macchetto, F. 1993, ApJ, 406, 451

\ref Martin, G.A., Fuhr, J.R., \& Wiese, W.L. 1988, J. Phys. Chem. Ref. Data,
17, Suppl. 3

\ref Mathews, W.G. \& Ferland, G.J. 1987, ApJ, 323, 456

\ref Moore, C.E. 1945, NSRDS-NBS 40

\ref Netze, H., \& Wills, B.J. 1983, ApJ, 275, 445

\ref Phinney, E.S. 1983, Ph.D. dissertation, Univ. of Canbridge

\ref Punsly, B., \& Coroniti, F.V. 1990, ApJ, 350, 518

\ref Schmidt, M. 1974, ApJ, 193, 509

\ref Wills, B.J., Netzer, H. \& Wills, D. 1985, ApJ, 288, 94

\newpage

\noindent{\bf Figure captions}

\bigskip

Fig.1 \ \ Normalized spectrum, from 3500 to 5250 \AA of IRAS 07598+6508. The
vertical scale gives the ratio of the observed flux (see text in \S 2).
In the
space above the spectrum we have marked the locations of all FeII lines,
excepting two, that originate from the six lowest odd-parity terms of FeII,
fall within the observed wavelength range of 3600-5200 \AA, and have values
of g$_u$A $\geq$ 10$^4$ s$^{-1}$. The multiplet number (Moore 1945) are also
listed. A line, at an observed wavelength of 4014 \AA (g$_u$A= 3.59 $\times$
10$^4$ s$^{-1}$), belonging to the transition from z$^4$D to a$^2$G, has not
been plotted because it does not have a multiplet number. Another line at
4752 \AA (g$_u$A = 1.07$\times$10$^4$ s$^{-1}$), belonging to the transition
from z$^4$P to b$^4$F (multiplet 39), has also not been plotted because of
insufficient room in that part of figure. We have also marked the Ca II H and
K doublet and, in the space below the spectrum, multiplets originating from
the two lowest odd-parity terms of TiII, and multiplets originating from the
15 lowest odd-parity terms of FeI.

\bigskip

Fig.2 \ \ Comparison of the observed spectra of IRAS 07598+6508 and PHL
1092.

\end{document}